\documentclass[11pt]{article}
\usepackage{epsfig}
\begin{document}
\vspace*{-3.5cm}
\noindent
DESY 00--169 \\
UG--FT--125/00 \\
hep-ph/0012305 \\
December 2000
\vspace{0.5cm}
\begin{center}
\begin{Large}
{\bf Top flavour-changing neutral coupling signals \\[0.2cm]
  at a linear collider}
\end{Large}

\vspace{0.5cm}
J. A. Aguilar--Saavedra \\
{\it Departamento de F\'{\i}sica Te\'{o}rica y del Cosmos \\
Universidad de Granada \\
E-18071 Granada, Spain} \\
\end{center}
\begin{abstract}
We present an analysis of the sensitivity of the TESLA $e^+ e^-$ collider to top
flavour-changing neutral couplings to the $Z$ boson and photon.
We consider the cases without beam
polarization, with only $e^-$ polarization and with $e^-$ and $e^+$
polarization, showing that the use of the latter substantially enhances the
sensitivity to discover or bound these vertices. For some of the couplings
the expected LHC limits could be improved up to an order of magnitude
for equal running times.
\end{abstract}


\section{Introduction}
\label{sec:1}
It is widely believed that the top quark, because of its large mass, will be a
sensitive probe into physics beyond the Standard Model (SM) \cite{papiro1}. In
particular, its couplings to the gauge and Higgs bosons may show deviations with
respect to the SM predictions. We will focus on flavour-changing neutral (FCN)
interactions involving the top, a light charge $2/3$ quark $q$ and a neutral
gauge boson $V=Z,\gamma$. In the SM the FCN couplings $Vtq$ vanish at tree-level
but can be generated at the one-loop level. However,
they are very suppressed by the GIM mechanism, because the
masses of the charge $-1/3$ quarks in the loop are small compared to the
scale involved. The calculation of the branching ratios for top decays mediated
by these FCN operators yields the SM predictions
$\mathrm{Br}(t \to Zc) = 1.3 \times 10^{-13}$,
$\mathrm{Br}(t \to \gamma c) = 4.5 \times 10^{-13}$ \cite{papiro12}, and smaller
values for the up quark. However, in many simple SM extensions these rates can
be orders of magnitude larger. For instance, in models with exotic quarks
$\mathrm{Br}(t \to Zq)$ can be of order $10^{-3}$ \cite{papiro13}.
Two Higgs doublet models allow for $\mathrm{Br}(t \to Zc) = 10^{-6}$,
$\mathrm{Br}(t \to \gamma c) = 10^{-7}$ \cite{papiro14}, and in $R$ 
parity-violating supersymmetric models one can have
$\mathrm{Br}(t \to Zc) = 10^{-4}$, $\mathrm{Br}(t \to \gamma c) = 10^{-5}$
\cite{papiro15}. Top FCN decays into a light Higgs boson and an up or charm
quark can also have similar or larger rates in these models
\cite{papiro15c,papiro14,papiro15b}.
Hence, top FCN couplings offer a good place to search for new
physics, which may manifest if these vertices are observed in
future colliders. At present the best limits on $Ztq$ couplings come from LEP 2,
$\mathrm{Br}(t \to Zq) \leq 0.07$ \cite{papiro25a,papiro25b}, and the best
limits on $\gamma tq$ couplings from Tevatron,
$\mathrm{Br}(t \to \gamma q) \leq 0.032$
\cite{papiro26}. They are very weak but will improve in the next years, first
with Tevatron Run II, and later with the next generation of colliders.

The CERN LHC will be a top factory. With a $t \bar t$ production cross-section
of 830 pb, at its 100 fb$^{-1}$ luminosity phase it will produce
$8.3 \times 10^7$ top-antitop pairs per year, and
$3 \times 10^7$ single tops plus antitops via other processes
\cite{papiro2,papiro3}. This makes LHC an excellent machine
for the investigation of the top quark properties. The search for FCN top
couplings can be carried out examining two different types of processes. On the
one hand, we can look for rare top decays $t \to Zq$ \cite{papiro4},
$t \to \gamma q$ \cite{papiro5}, $t \to gq$ \cite{papiro6} or
$t \to Hq$ \cite{papiro7} of the tops or antitops produced in the SM process
$gg,q \bar q \to t \bar t$. On the other hand, one can search for single top
production via an anomalous effective vertex: $Zt$ and $\gamma t$ production
\cite{papiro8}, the production of a single top quark \cite{papiro9},
and $Ht$ production \cite{papiro7}. In these cases
the top quark is assumed to decay in the SM dominant mode $t \to Wb$.

The TESLA $e^+ e^-$ collider with a centre of mass (CM) energy $\sqrt s=500$
GeV has a tree-level
$t \bar t$ production cross-section of 0.52 pb, and produces only
$1.56 \times 10^5$ top-antitop pairs per year with an integrated luminosity of
300 fb$^{-1}$. However, $e^+ e^-$
colliders are cleaner than hadron colliders. For instance, the signal to
background ratio $S/B$ for rare top decays can be 7 times larger in TESLA than
in LHC. But the sensitivity to rare top decays is given in the Gaussian
statistics limit by the ratio $S/\sqrt B$, and the larger LHC cross-sections
make difficult for TESLA to compete with it in the search for anomalous top
couplings.

Here we show that for single top production \cite{papiro16}
the use of beam polarization in TESLA substantially
enhances the sensitivity to discover or bound top anomalous FCN couplings and
is equivalent to an increase in the luminosity by a factor of $6-7$. This allows
TESLA to improve some of the expected LHC limits up to an order of magnitude.
We consider the planned CM energies of 500 and 800 GeV, and for both we analyse
three cases: without beam polarization, with $80\%$ $e^-$ polarization, and
with $80\%$ $e^-$, $45\%$ $e^+$ polarization.
We also study rare top decays in the
processes $e^+ e^- \to t \bar t$, with subsequent decay $\bar t \to V \bar q$
(or $t \to V q$). Single top production and top decay processes are
complementary: although
single top production is more sensitive to top anomalous couplings, top
decays can help to disentangle the type of anomalous coupling involved ($Ztq$
or $\gamma tq$) if a positive signal is discovered.

\section{Single top production}
\label{sec:2}

In order to describe the FCN couplings among the top, a light quark $q$ and a
$Z$ boson or a photon $A$ we use the Lagrangian
\begin{eqnarray}
-{\mathcal L} & = & \frac{g_W}{2 c_W} \, X_{tq} \, \bar t \gamma_\mu
 (x_{tq}^L P_L +  x_{tq}^R P_R) q Z^\mu 
+ \frac{g_W}{2 c_W} \, \kappa_{tq}\, \bar t (\kappa_{tq}^{v}- \kappa_{tq}^{a}
\gamma_5) \frac{i \sigma_{\mu \nu} q^\nu}{m_t} q Z^\mu  \nonumber \\
& &  + e \, \lambda_{tq}\, \bar t (\lambda_{tq}^{v}- \lambda_{tq}^{a} \gamma_5)
\frac{i \sigma_{\mu \nu} q^\nu}{m_t} q A^\mu  \,, \label{ec:1}
\end{eqnarray}
where $P_{R,L}=(1 \pm \gamma_5)/2$. The chirality-dependent parts are normalized
to $(x_{tq}^L)^2+(x_{tq}^R)^2=1$, $(\kappa_{tq}^{v})^2+(\kappa_{tq}^{a})^2=1$,
$(\lambda_{tq}^{v})^2+(\lambda_{tq}^{a})^2=1$. This effective Lagrangian
contains $\gamma_\mu$ terms of dimension 4 and $\sigma_{\mu \nu}$ terms of
dimension 5. The couplings are constants corresponding to the first terms in
the expansion in momenta. The $\sigma_{\mu \nu}$ terms are the only ones allowed
by the unbroken gauge symmetry, ${\mathrm SU(3)}_c \times {\mathrm U(1)}_Q$. Due
to their extra momentum factor they grow with the energy and make single top
production at TESLA the best process to measure them.

\begin{figure*}[htb]
\begin{center}
\mbox{\epsfig{file=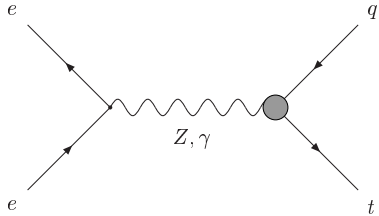,width=5cm}}
\end{center}
\caption{Feynman diagrams for $e^+ e^- \to t\bar q$ via $Ztq$ or $\gamma tq$
FCN couplings. The top quark is off-shell and has the SM decay.
\label{fig:feyn1} }
\end{figure*}

For single top production we study the process $e^+ e^- \to t \bar q$ mediated
by $Ztq$ or $\gamma tq$ anomalous couplings as shown in Fig.~\ref{fig:feyn1},
followed by top decay $t \to W^+ b \to l^+\nu b$, with $l=e,\mu$. We will
only take one type of anomalous coupling different from zero at the same time,
and we evaluate three signals: ({\em i\/}) with $Ztq$ $\gamma_\mu$ couplings,
setting $x_{tq}^L = x_{tq}^R$ for definiteness; ({\em ii\/}) with $Ztq$
$\sigma_{\mu \nu}$ couplings, taking $\kappa_{tq}^v=1$, $\kappa_{tq}^a=0$;
({\em iii\/}) with $\gamma tq$ couplings, taking $\lambda_{tq}^v=1$,
$\lambda_{tq}^a=0$. However,
if a positive signal is discovered, it may be difficult to distinguish only from
this process whether the anomalous coupling involves the $Z$ boson, the photon
or both. On the other hand, in principle it could be possible to have a
fine-tuned cancellation between $Z$ and $\gamma$ contributions that led to a
suppression of this signal.

We calculate the matrix element for
$e^+ e^- \to t \bar q \to W^+ b \bar q \to l^+ \nu b \bar q$
using HELAS \cite{papiro18}
and introducing a new HELAS-like subroutine {\tt IOV2XX} to compute the
non-renormalizable $\sigma_{\mu \nu}$ vertices. This new routine has been
checked comparing with calculations done
by hand. In all cases we sum $t \bar q + \bar t q$ production but we sometimes
refer to the signals as $t\bar q$ for simplicity.
The signal cross-sections depend slightly on the chirality of the anomalous
couplings. For a CM energy of 500 GeV and the
three polarization options discussed the cross-sections for $\gamma_\mu$
and $\sigma_{\mu \nu}$ couplings have a maximum variation of $1.2\%$ for other
chirality choices.

The background for the $t \bar q$ signal is given by $W^+ q \bar q'$ production
with $W^+$ decay to electrons and muons. The leading contribution to this
process is $W^+ W^-$ production with $W^-$ hadronic decay, but it is crucial for
the correct evaluation of the background after kinematical cuts to take into
account the 7 interfering Feynman diagrams for $e^+ e^- \to W^+ q \bar q'$.
Taking all the interfering diagrams for $e^+ e^- \to l^+ \nu q \bar q'$ into
account does not give an appreciable difference in the cross-section.
This background is evaluated using MadGraph
\cite{papiro19} and modifying the code to include the $W^+$ decay.

To simulate the calorimeter energy resolution we perform a Gaussian smearing of
the charged lepton ($l$), photon ($\gamma$) and jet ($j$) energies using a
calorimeter resolution \cite{papiro20} of
\begin{equation}
\frac{\Delta E^{l,\gamma}}{E^{l,\gamma}} = \frac{10\%}{\sqrt{E^{l,\gamma}}}
\oplus 1\% \,, ~~~~
\frac{\Delta E^{j}}{E^{j}} = \frac{50\%}{\sqrt{E^j}} \oplus 4\% \,,
\end{equation}
where the energies are in GeV and the two terms are added in quadrature. For
simplicity we assume that the energy smearing for muons is the same as for
electrons. Note that more optimistic resolutions would improve our results. We
then apply detector cuts on transverse momenta, $p_T \geq 10 ~\mathrm{GeV}$, and
pseudorapidities, $|\eta| \leq 2.5$ (this corresponds to polar angles
$10^\circ \leq \theta \leq 170^\circ$).
We reject the events in which the jets and/or leptons are not
isolated, requiring that the distances in $(\eta,\phi)$ space
satisfy $\Delta R \geq 0.4$. We do not require specific trigger conditions,
and we assume that the presence of high $p_T$ charged leptons will suffice.

\begin{figure}[htb]
\begin{center}
\mbox{\epsfig{file=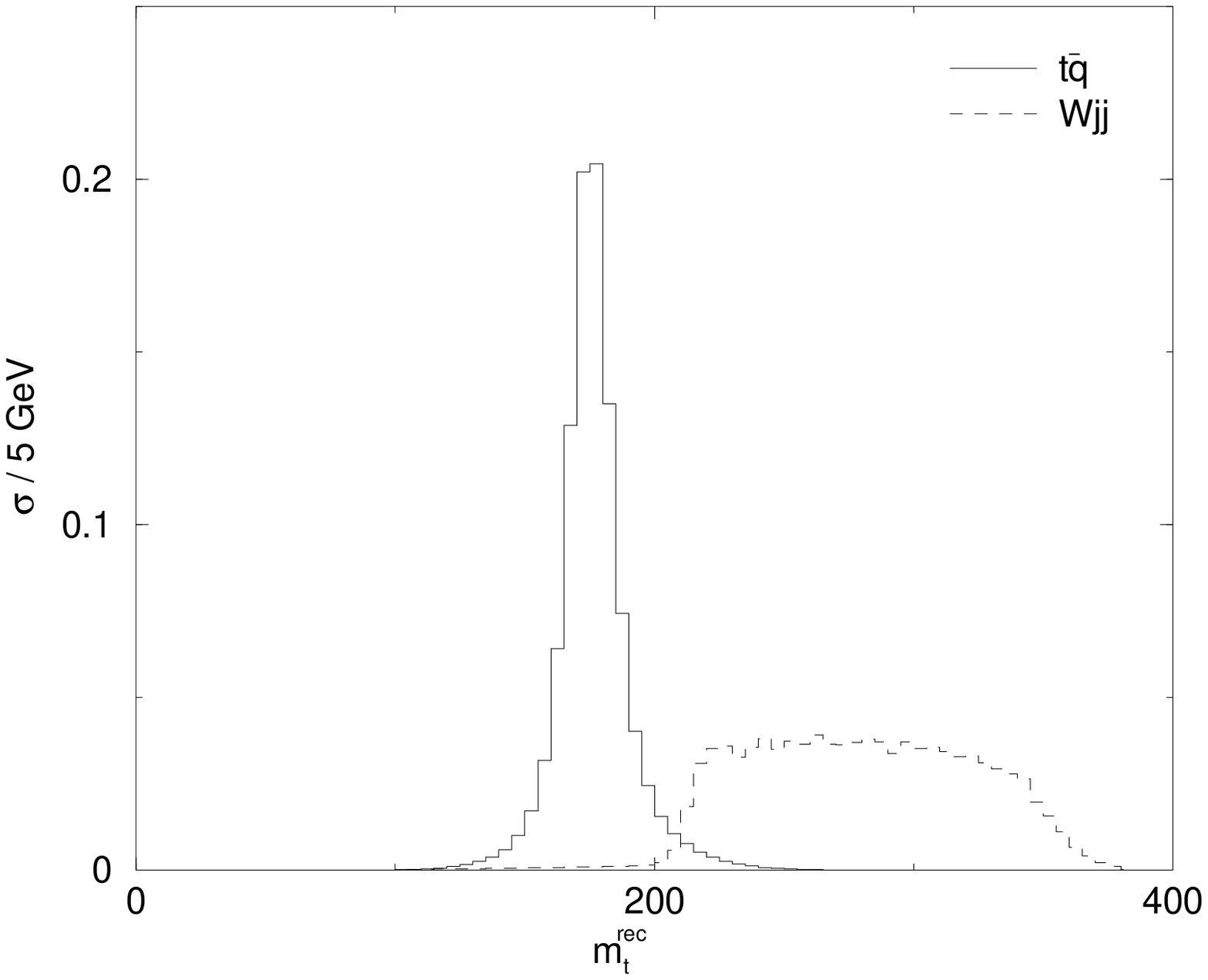,width=8cm}}
\end{center}
\caption{Reconstructed top mass $m_t^\mathrm{rec}$ distribution before
kinematical cuts for the three $t \bar q$ signals and $W^+ jj$ background at a
CM energy of 500 GeV, without beam polarization. The cross-sections are
normalized to unity.
\label{fig:tc-mt}}
\end{figure}

\begin{figure}[htb]
\begin{center}
\mbox{\epsfig{file=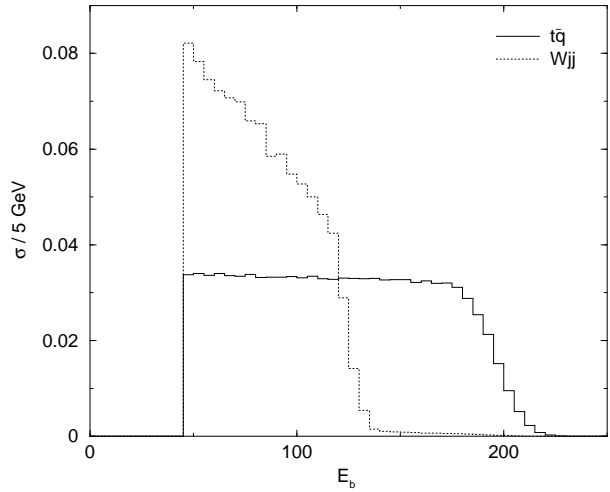,width=8cm}}
\end{center}
\caption{$E_b$ distribution before kinematical cuts for the three $t \bar q$
signals and $W^+ jj$ background at a CM energy of 500 GeV, without beam
polarization. The cross-sections are normalized to unity.
\label{fig:tc-eb}}
\end{figure}

The signal is reconstructed as follows. The neutrino momentum $p_\nu$ can be
identified with the missing momentum of the event. The longitudinal missing
momentum can also be used, and $p_\nu$ is reconstructed without any ambiguity.
The $W^+$ momentum is then the sum of the momenta of the charged lepton and the
neutrino. For the signal, the invariant mass of the $W^+$
and one of the jets, $m_t^\mathrm{rec}$, is consistent with the top mass,
and the other jet has an energy $E_q$ around
$E_q^0 \equiv (s-m_t^2)/(2 \sqrt s)$. Of the
two possible pairings, we choose the one minimizing
$(m_t^\mathrm{rec}-m_t)^2+(E_q-E_q^0)^2$. The kinematical distributions of
$m_t^\mathrm{rec}$ for the signal and the background at a CM
energy of 500 GeV without beam polarization are plotted in
Fig.~\ref{fig:tc-mt}. Then we require a $b$ tag on the jet associated to
the decay of the top quark to reduce the background. To ensure a high charm
rejection we apply more strict kinematical cuts on this jet,
$|\eta_b| \leq 2$ (polar angle $15^\circ \leq \theta_b \leq
165^\circ$) and energy $E_b \geq 45$ GeV. We assume a $b$ tagging efficiency of
$60\%$, and mistagging rates of $5\%$ for charm and $0.5\%$ for lighter quarks
\cite{papiro21}. The $E_b$
kinematical distributions are shown in Fig.~\ref{fig:tc-eb}.

\begin{figure}[htb]
\begin{center}
\mbox{\epsfig{file=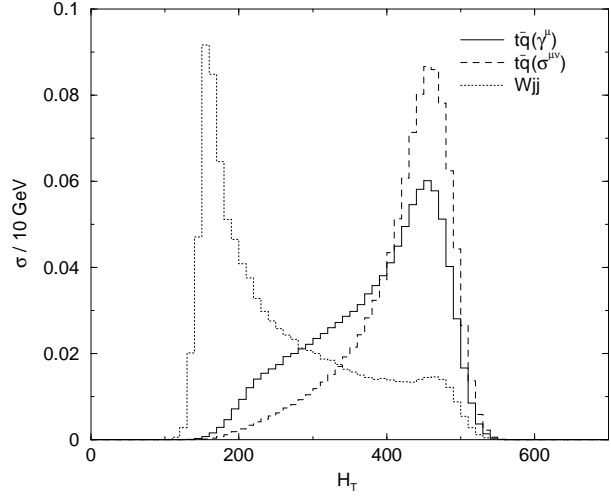,width=8cm}}
\end{center}
\caption{Total transverse energy $H_T$ distribution before kinematical cuts for
the three $t \bar q$ signals and $W^+ jj$ background at a CM energy of 500 GeV,
without beam polarization. The cross-sections are normalized to unity.
\label{fig:tc-ht}}
\end{figure}

\begin{figure}[htb]
\begin{center}
\mbox{\epsfig{file=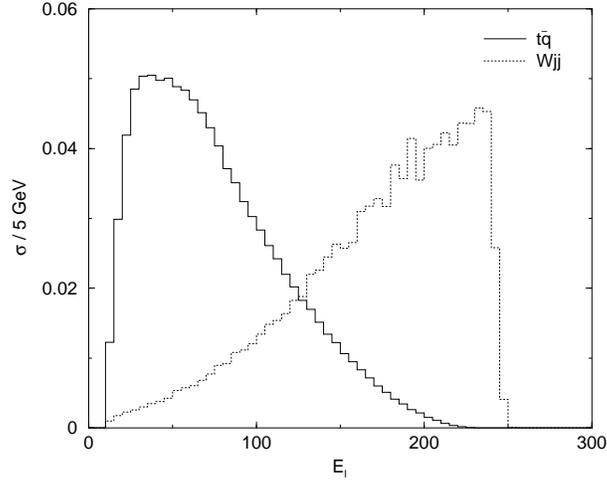,width=8cm}}
\end{center}
\caption{Charged lepton energy $E_l$ distribution before kinematical cuts for
the three $t \bar q$ signals and $W^+ jj$ background at a CM energy of 500 GeV,
without beam polarization. The cross-sections are normalized to unity.
\label{fig:tc-el}}
\end{figure}

Another interesting variable to distinguish the signal from the background
is the two-jet invariant mass $M_{W^-}^\mathrm{rec}$. The $W^+jj$ background is
dominated by $W^+ W^-$ production with $W^- \to jj$, and the
$M_{W^-}^\mathrm{rec}$ distribution peaks around $M_W$. A veto cut on
$M_{W^-}^\mathrm{rec}$ can eliminate a
large fraction of the background but makes compulsory to calculate correctly the
cross-section to include all the diagrams for $e^+ e^- \to W^+ q \bar q'$. Also
of interest are the total transverse energy $H_T$ and the charged lepton energy
$E_l$ in Figs.~\ref{fig:tc-ht} and \ref{fig:tc-el}. The kinematical
distributions with polarized beams are very similar except the $H_T$
distribution. In this case polarization decreases the peak of the background
around $H_T=200$.

To enhance the signal significance we perform kinematical cuts on these
variables. However, we find that the veto cut on $M_{W^-}^\mathrm{rec}$ is
unnecessary in single top production since the requirement $E_b>45$ GeV and the
kinematical cut on $m_t^\mathrm{rec}$ practically eliminate the peak in the
$M_{W^-}^\mathrm{rec}$
distribution. A cut on $E_q$ is unnecessary because this
variable is kinematically related to $m_t^\mathrm{rec}$, and we prefer to apply
a cut on $m_t^\mathrm{rec}$ to show the presence of a top quark in the signal.
For simplicity, we choose to apply the same cuts for the three
signals and the three polarization options, but different for CM energies of 500
and 800 GeV. We choose the cuts trying to maintain the independence of the
cross-section on the chirality of the coupling. Obviously, our results could be
improved modifying the cuts for each type of coupling and each polarization
option. 
Before discussing the results it is convenient to outline the procedure
used to obtain the limits on the anomalous couplings. The correct statistical
treatment of signals and backgrounds is specially necessary in our study since
the backgrounds are very small even for high integrated luminosities.

Assuming that no signal is observed after the experiment is done, {\em i.e.}
the number of observed events $n_0$ equals the expected background $n_b$, we
derive $95\%$ confidence level (CL) upper bounds on the number of events
expected $n_s$. We use the Feldman-Cousins construction for the confidence
intervals of a Poisson variable \cite{papiro22} evaluated with the PCI package
\cite{papiro23}.

On the other hand, we can obtain the smallest value of $n_s$ such that a
positive signal is expected to be observed with $3\,\sigma$ significance,
assuming that the number of observed events for $3\,\sigma$ `evidence' $n_e$
equals $n_s+n_b$. For a large number of background events, the Poisson
probability distribution can be approximated by a Gaussian of mean $n_b$ and
standard deviation $\sqrt n_b$. The requirement of $3\,\sigma$ significance is
then simply $n_s/\sqrt n_b \geq 3$. However, this is seldom the case for our
study, where the backgrounds are very small. In such case, we use the estimator
based on the $\mathcal{P}$ number (see for example \cite{papiro24}). The number
$\mathcal{P}(n)$ is defined as the probability of the background to fluctuate
and give $n$ or more observed events. $n_e$ is then defined as the smallest
value of $n$ such that $1-\mathcal{P}(n) \geq 99.73\%$, corresponding to three
Gaussian standard deviations.

The limits on the number of signal events obtained in this way can be
translated into limits on the anomalous coupling constants, and expressed
in terms of top decay branching ratios taking $\Gamma_t = 1.56$ GeV. We now
discuss the results for 500 GeV and 800 GeV in turn.

\subsection{Limits for $\sqrt s = 500$ GeV}

The kinematical cuts for 500 GeV are collected in Table~\ref{tab:tc1}, and the
cross-sections before and after cuts in
Table~\ref{tab:tc2}. We normalize the signals to $X_{tq} = 0.06$,
$\kappa_{tq} = 0.02$, $\lambda_{tq} = 0.02$ and sum $t \bar q+\bar t q$
production. For different chiralities of the anomalous couplings the
cross-sections
after cuts differ at most $7\%$ for $\gamma_\mu$ couplings and $5\%$ for
$\sigma_{\mu \nu}$ couplings.

\begin{table}[htb]
\begin{center}
\begin{tabular}{cc}
Variable & Cut \\
$m_t^\mathrm{rec}$ & $160-190$ \\
$H_T$ & $>220$ \\
$E_l$ & $<160$ 
\end{tabular}
\caption{Kinematical cuts for the three $t \bar q$ signals and the three
polarization options at a CM energy of 500 GeV. The masses and the energies are
in GeV.
\label{tab:tc1}}
\end{center}
\end{table}

\begin{table}[htb]
\begin{center}
\begin{tabular}{ccccccc}
& \multicolumn{2}{c}{No pol.} & \multicolumn{2}{c}{Pol. $e^-$}
& \multicolumn{2}{c}{Pol. $e^-$ $e^+$} \\
& before & after & before & after & before & after \\[-0.2cm]
& cuts & cuts & cuts & cuts & cuts & cuts \\
$t \bar q+\bar tq$ $(Z,\gamma_\mu)$
           & 0.183 & 0.137 & 0.162 & 0.121 & 0.215 & 0.161 \\
$t \bar q+\bar tq$ $(Z,\sigma_{\mu \nu})$
           & 0.199 & 0.153 & 0.176 & 0.135 & 0.234 & 0.179 \\
$t \bar q+\bar tq$ $(\gamma)$
           & 0.375 & 0.288 & 0.375 & 0.287 & 0.510 & 0.391 \\
$W^\pm jj$ & 19.5  & 0.0734 & 4.06  & 0.0154 & 2.40 & 0.0092
\end{tabular}
\caption{Cross-sections (in fb) before and after the kinematical cuts in
Table~\ref{tab:tc1} for the three $t \bar q$ signals and their background at a
CM energy of 500 GeV, for the three polarization options. We include $b$ tagging
efficiencies and use $X_{tq} = 0.06$, $\kappa_{tq} = 0.02$,
$\lambda_{tq} = 0.02$.
\label{tab:tc2}}
\end{center}
\end{table}

Polarization is very useful to improve the limits from single top production.
In Table~\ref{tab:tc2} we notice that the use of
$80\%$ $e^-$ polarization decreases the background by a factor of 5 while
keeping $90\%$ of the signal, and additional $e^+$ polarization of $45\%$
decreases the background by a factor of 8 and increases the signal $20\%$
with respect to the values without polarization. The improvement is clearly
seen in the limits of
Table~\ref{tab:tc4}, obtained for an integrated luminosity of 300 fb$^{-1}$.
$e^-$, $e^+$ polarization improves the $3\,\sigma$ discovery limits by factors
of $2.8-3.2$. The integrated luminosity required to obtain the
same limits without using polarization would be 2100 fb$^{-1}$.

\begin{table}[htb]
\begin{center}
\begin{small}
\begin{tabular}{lcccccc}
& \multicolumn{2}{c}{No pol.} & \multicolumn{2}{c}{Pol. $e^-$}
& \multicolumn{2}{c}{Pol. $e^-$ $e^+$} \\
& $95\%$ & $3\,\sigma$ & $95\%$ & $3\,\sigma$ & $95\%$ & $3\,\sigma$ \\
$\mathrm{Br}(t \to Zq)$ $(\gamma_\mu)$ &
  $4.4 \times 10^{-4}$ & $6.1 \times 10^{-4}$ &
  $3.1 \times 10^{-4}$ & $3.9 \times 10^{-4}$ &
  $1.9 \times 10^{-4}$ & $2.2 \times 10^{-4}$ \\
$\mathrm{Br}(t \to Zq)$ $(\sigma_{\mu \nu})$ &
  $3.5 \times 10^{-5}$ & $4.8 \times 10^{-5}$ &
  $2.4 \times 10^{-5}$ & $3.1 \times 10^{-5}$ &
  $1.5 \times 10^{-5}$ & $1.7 \times 10^{-5}$ \\
$\mathrm{Br}(t \to \gamma q)$ &
  $2.2 \times 10^{-5}$ & $3.0 \times 10^{-5}$ &
  $1.3 \times 10^{-5}$ & $1.7 \times 10^{-5}$ &
  $8.2 \times 10^{-6}$ & $9.3 \times 10^{-6}$
\end{tabular}
\end{small}
\caption{Limits on top FCN decay branching ratios obtained from single top
production at 500 GeV with a luminosity of 300 fb$^{-1}$ for the three
polarization options.
\label{tab:tc4}}
\end{center}
\end{table}

\subsection{Limits for $\sqrt s = 800$ GeV}

We write the kinematical cuts for 800 GeV in Table~\ref{tab:tc5}, and the
signal cross-sections before and after cuts in Table~\ref{tab:tc6}.
The cross-sections for $t\bar q$ production mediated by non-renormalizable
couplings do not decrease raising the CM energy from 500 to 800 GeV,
whereas the background
decreases to less than one half.
This improves the sensitivity for $\sigma_{\mu \nu}$ couplings with respect to
500 GeV. Unfortunately, the signal with $\gamma_\mu$ couplings also decreases,
and thus the results are worse in this case.

\begin{table}[htb]
\begin{center}
\begin{tabular}{ccc}
Variable & Cut \\
$m_t^\mathrm{rec}$ & $160-190$ \\
$H_T$ & $>450$ \\
$E_l$ & $<300$
\end{tabular}
\caption{Kinematical cuts for the three $t \bar q$ signals and the three
polarization options at a CM energy of 800 GeV. The masses and the energies are
in GeV.
\label{tab:tc5}}
\end{center}
\end{table}

\begin{table}[htb]
\begin{center}
\begin{tabular}{lcccccc}
& \multicolumn{2}{c}{No pol.} & \multicolumn{2}{c}{Pol. $e^-$}
& \multicolumn{2}{c}{Pol. $e^-$ $e^+$} \\
& before & after & before & after & before & after \\[-0.2cm]
& cuts & cuts & cuts & cuts & cuts & cuts \\
$t \bar q+\bar tq$ $(Z,\gamma_\mu)$
           & 0.0776 & 0.0498 & 0.0684 & 0.0440 & 0.0912 & 0.0586 \\
$t \bar q+\bar tq$ $(Z,\sigma_{\mu \nu})$
           & 0.198 & 0.149 & 0.175 & 0.132 & 0.233 & 0.175 \\
$t \bar q+\bar tq$ $(\gamma)$
           & 0.389 & 0.293 & 0.389 & 0.293 & 0.528 & 0.398 \\
$W^\pm jj$ & 8.45   & 0.0125 & 1.75   & 0.0028 & 1.03   & 0.0018
\end{tabular}
\caption{Cross-sections (in fb) before and after the kinematical cuts in
Table~\ref{tab:tc5} for the three $t \bar q$ signals and their background at a
CM energy of 800 GeV, for the three polarization options. We include $b$ tagging
efficiencies and use $X_{tq} = 0.06$, $\kappa_{tq} = 0.02$,
$\lambda_{tq} = 0.02$.
\label{tab:tc6}}
\end{center}
\end{table}

For equal luminosities, an $e^+ e^-$ collider with a CM energy of 800 GeV is
sensitive to top rare decays mediated by $\sigma_{\mu \nu}$ vertices with
branching ratios $1.5-2$ times smaller than one with 500 GeV. Of course, the
higher luminosity at 800 GeV has also to be taken into account, and then this
energy is best suited to perform searches for these vertices.
Again we observe the usefulness of polarization: the use of $e^-$ polarization
reduces the background 5 times and the use of $e^+$ polarization as well
reduces it 8 times. In Table~\ref{tab:tc8}
we gather the limits for an integrated luminosity of 500 fb$^{-1}$.
$e^-$, $e^+$ polarization improves the $3\,\sigma$ discovery limits
by factors of $2.5-2.9$.
The luminosity necessary to obtain the same limits without
the use of polarization would be 3000 fb$^{-1}$.

\begin{table}[htb]
\begin{center}
\begin{small}
\begin{tabular}{lcccccc}
& \multicolumn{2}{c}{No pol.} & \multicolumn{2}{c}{Pol. $e^-$}
& \multicolumn{2}{c}{Pol. $e^-$ $e^+$} \\
& $95\%$ & $3\,\sigma$ & $95\%$ & $3\,\sigma$ & $95\%$ & $3\,\sigma$ \\
$\mathrm{Br}(t \to Zq)$ $(\gamma_\mu)$ &
  $4.4 \times 10^{-4}$ & $5.9 \times 10^{-4}$ &
  $2.9 \times 10^{-4}$ & $4.3 \times 10^{-4}$ &
  $2.4 \times 10^{-4}$ & $2.3 \times 10^{-4}$ \\
$\mathrm{Br}(t \to Zq)$ $(\sigma_{\mu \nu})$ &
  $1.3 \times 10^{-5}$ & $1.7 \times 10^{-5}$ &
  $8.6 \times 10^{-6}$ & $1.3 \times 10^{-5}$ &
  $6.2 \times 10^{-6}$ & $7.0 \times 10^{-6}$ \\
$\mathrm{Br}(t \to \gamma q)$ &
  $7.8 \times 10^{-6}$ & $1.0 \times 10^{-5}$ &
  $4.5 \times 10^{-6}$ & $6.7 \times 10^{-6}$ &
  $3.7 \times 10^{-6}$ & $3.6 \times 10^{-6}$
\end{tabular}
\caption{Limits on top FCN decay branching ratios obtained from single top
production at 800 GeV with a luminosity of 500 fb$^{-1}$ for the three
polarization options.
\label{tab:tc8}}
\end{small}
\end{center}
\end{table}

\section{Top decays}
\label{sec:3}

For top decays we study the SM process $e^+ e^- \to t \bar t$, followed by
antitop decay mediated by an anomalous $Ztq$ or $\gamma tq$ coupling.
This gives the signals $t \bar qZ$ and $t \bar q\gamma$, and the different
final states distinguish $Ztq$ and $\gamma tq$ couplings. The
top is assumed to decay via $t \to W^+b \to l^+\nu b$, with $l=e,\mu$. For the
$t \bar qZ$ signal we only consider the $Z$ boson decays to electrons and muons.

The cross-sections for the $t\bar qV$ signals are smaller than for single top
production for equal values of the FCN coupling parameters. The reasons are:
({\em i\/}) $t \bar q$ production is enhanced by the $q^\nu$ factor of the
$\sigma_{\mu \nu}$ vertex, when present, whereas $t \bar qV$ is not;
({\em ii\/}) the final state cross-section for $t \bar qZ$ includes the partial
width $\mathrm{Br}(Z \to l'^+ l'^-) = 0.067$ ; ({\em iii\/}) phase space for the
production of a $t \bar t$ pair is smaller than for $t \bar q$. However,
top decay signals are cleaner than single top production. This can be
understood since the top decay signals $W^+ bjV$ have the
enhancement over their background $W^+ jjV$ of two on-shell particles, the top
and the antitop, whereas single top production has only the enhancement due to
the top on-shell and the $\sigma_{\mu \nu}$ coupling if that is the case.

For $t \bar qZ$ and $t \bar q\gamma$ production we calculate the matrix
elements
$e^+ e^- \to t \bar t \to W^+ b \bar q Z \to l^+ \nu b \bar q l'^+ l'^-$ and
$e^+ e^- \to t \bar t \to W^+ b \bar q \gamma \to l^+ \nu b \bar q \gamma$,
respectively, using HELAS as for $t \bar q$ production.
For the $t \bar qV$ signals there is an additional contribution from $t \bar q$
production plus radiative emission of a $Z$ boson or a photon. This correction
is suppressed because it does not have the enhancement due to the $\bar t$
on-shell, and is even smaller after the kinematical cuts for the signal
reconstruction. We assume only one type of anomalous coupling different from
zero at the same time, and give our results for the same chiralities
used before. We check that for other chirality choices the
differences are of order $0.1\%$
for the three polarization options, before and after kinematical cuts.
The backgrounds for the $t \bar qZ$ and $t \bar q\gamma$ signals are
$W^+ q \bar q'Z$ and $W^+ q \bar q'\gamma$, with 46 and 44 diagrams,
respectively. They are calculated with MadGraph.

After energy smearing and detector cuts,
the $t \bar q \gamma$ signal can be reconstructed in a similar way as $t \bar
q$. The $W^+$ momentum is the charged lepton momentum plus the missing momentum.
The invariant mass of the $W^+$ and one of the jets, $m_t^\mathrm{rec}$,
is consistent with the top mass, and the invariant mass of the photon and the
other jet, $m_{\bar t}^\mathrm{rec}$, is also consistent with $m_t$. Of the two
possible assignments, we choose the one minimizing
$(m_t^\mathrm{rec}-m_t)^2+(m_{\bar t}^\mathrm{rec}-m_t)^2$ and require a
$b$ tag on the jet that corresponds to the top quark. The reconstructed $W^-$
mass $M_{W^-}^\mathrm{rec}$ is defined as the invariant mass of the two jets as
before.

The reconstruction of the $t \bar q Z$ signal is slightly more
complicated. Of the
two positively charged leptons, one results from the $W^+$ decay and it has with
the neutrino (reconstructed from the missing momentum)  an invariant mass
$M_{W^+}^\mathrm{rec}$ consistent with $M_W$. The other one and the negative
charge lepton have an invariant mass $M_Z^\mathrm{rec}$ close to $M_Z$. If the
two positive leptons have different flavours the assignment is straightforward,
but if they have not we choose the pairing that minimizes
$(M_{W^+}^\mathrm{rec}-M_W)^2+(M_Z^\mathrm{rec}-M_Z)^2$. Then, we
reconstruct the top and antitop masses as for the $t \bar q \gamma$ signal
replacing the photon momentum by the $Z$ momentum. The $W^-$ reconstruction
for the background is also similar. 

In our analysis we find that all the signal cross-sections, including those with
$\sigma_{\mu \nu}$ vertices, decrease raising the CM energy from 500 to 800 GeV,
and for the latter the limits obtained are worse even after taking into account
the increased luminosity. Hence we will discuss only the results for top decays
at 500 GeV. We write the kinematical cuts in Table~\ref{tab:tcv1}. The cut on
$m_{\bar t}^\mathrm{rec}$ is more strict than the cut on
$m_t^\mathrm{rec}$ because the reconstruction of the antitop mass is
better. The cuts for $t\bar qZ$ are looser because the background is much
smaller. Contrarily to $t\bar q$ production, the veto cuts on  
$M_{W^-}^\mathrm{rec}$ are not redundant.
The cross-sections before and after cuts are collected in
Table~\ref{tab:tcv2}. Note that we normalize the signal to $X_{tq} = 0.2$,
$\kappa_{tq} = 0.2$ $\lambda_{tq} = 0.04$ because the
signal cross-sections are smaller.

\begin{table}[htb]
\begin{center}
\begin{tabular}{cccc}
Variable & $t\bar q Z$ cut & $t \bar q\gamma$ cut \\
$m_t^\mathrm{rec}$ & $130-220$ & $150-200$ \\
$m_{\bar t}^\mathrm{rec}$ & $150-200$ & $160-190$ \\
$M_{W^-}^\mathrm{rec}$ & $<70$ or $>90$ & $<65$ or $>95$ 
\end{tabular}
\caption{Kinematical cuts for the  $t\bar qV$ signals and
the three polarization options at a CM energy of 500 GeV. The masses are in GeV.
\label{tab:tcv1}}
\end{center}
\end{table}

\begin{table}[htb]
\begin{center}
\begin{tabular}{lcccccc}
& \multicolumn{2}{c}{No pol.} & \multicolumn{2}{c}{Pol. $e^-$}
& \multicolumn{2}{c}{Pol. $e^-$ $e^+$} \\
& before & after & before & after & before & after \\[-0.2cm]
& cuts & cuts & cuts & cuts & cuts & cuts \\
$t \bar qZ +\bar tqZ$ $(\gamma_\mu)$
            & 0.114  & 0.105  & 0.0784 & 0.0720 & 0.0995 & 0.0912 \\
$t \bar qZ +\bar tqZ$ $(\sigma_{\mu \nu})$
            & 0.0877 & 0.0809 & 0.0604 & 0.0555 & 0.0766 & 0.0703 \\
$t \bar q \gamma +\bar tq \gamma$
                 & 0.0745 & 0.0631 & 0.0515 & 0.0429 & 0.0653 & 0.0543 \\
$W^\pm jjZ$ & 0.0059 & $1.0 \times 10^{-4}$ & 0.0013 & $2.4 \times 10^{-5}$ &
$8.9 \times 10^{-4}$ & $1.6 \times 10^{-5}$ \\
$W^\pm jj\gamma$ & 0.639  & 0.0014 & 0.144  & $3.1 \times 10^{-4}$ & 0.0956 &
 $2.0 \times 10^{-4}$
\end{tabular}
\caption{Cross-sections (in fb) before and after the kinematical cuts in
Table~\ref{tab:tcv1} for the  $t \bar qV$ signals and backgrounds at a CM
energy of 500 GeV, for the three polarization options. We include $b$ tagging
efficiencies and use $X_{tq} = 0.2$, $\kappa_{tq} = 0.2$, $\lambda_{tq} = 0.04$.
\label{tab:tcv2}}
\end{center}
\end{table}

For top decay signals the use of polarization is not as useful as for single
top production. Although it reduces the $W^+ jjV$ cross-sections up to a factor
of 7, these backgrounds are
already very small for unpolarized beams, and the luminosities required to
glimpse the potential improvement would exceed 1000 fb$^{-1}$. In addition,
the signal cross-sections decrease $10-20\%$, in contrast to single top
production, and the limits
obtained are in some cases worse (see Table~\ref{tab:tcv4}).

\begin{table}[htb]
\begin{center}
\begin{small}
\begin{tabular}{lcccccc}
& \multicolumn{2}{c}{No pol.} & \multicolumn{2}{c}{Pol. $e^-$}
& \multicolumn{2}{c}{Pol. $e^-$ $e^+$} \\
& $95\%$ & $3\,\sigma$ & $95\%$ & $3\,\sigma$ & $95\%$ & $3\,\sigma$ \\
$\mathrm{Br}(t \to Zq)$ $(\gamma_\mu)$ &
  $1.8 \times 10^{-3}$ & $1.2 \times 10^{-3}$ &
  $2.7 \times 10^{-3}$ & $1.7 \times 10^{-3}$ &
  $2.1 \times 10^{-3}$ & $1.4 \times 10^{-3}$ \\
$\mathrm{Br}(t \to Zq)$ $(\sigma_{\mu \nu})$ &
  $1.9 \times 10^{-3}$ & $1.2 \times 10^{-3}$ &
  $2.8 \times 10^{-3}$ & $1.8 \times 10^{-3}$ &
  $2.2 \times 10^{-3}$ & $1.4 \times 10^{-3}$ \\
$\mathrm{Br}(t \to \gamma q)$ &
  $9.9 \times 10^{-5}$ & $1.3 \times 10^{-4}$ &
  $1.6 \times 10^{-4}$ & $1.6 \times 10^{-4}$ &
  $1.3 \times 10^{-4}$ & $8.3 \times 10^{-5}$ 
\end{tabular}
\end{small}
\caption{Limits on top FCN decay branching ratios obtained from the $t \bar qV$
signals at 500 GeV with a luminosity of 300 fb$^{-1}$ for the three
polarization options.
\label{tab:tcv4}}
\end{center}
\end{table}

The limits from the $t \bar qV$ signals are in all cases worse than those
obtained from single top production, especially for $Ztq$ couplings.
In fact, these processes would only be useful if a
FCN top decay is detected with $\mathrm{Br}(t \to Zq) \sim 10^{-3}$
or $\mathrm{Br}(t \to \gamma q) \sim 10^{-4}$. In such
case, they would help to determine the nature of the top anomalous coupling.
Besides, it is interesting to notice that the limits for $Ztq$ $\gamma_\mu$ and
$\sigma_{\mu \nu}$ couplings are remarkably similar, what confirms that top
decays are not sensitive to the $q^\nu$ factor of the $\sigma_{\mu \nu}$ vertex.

\section{Comparison with LHC}
\label{sec:4}

We have proved that, despite the relatively small single top and top pair
production cross sections, TESLA is able to observe top FCN vertices
corresponding to very small top decay branching ratios, or set competitive
bounds on them if they are not observed. To obtain these results,
beam polarization is essential to reduce the backgrounds. 
We now compare the best limits on anomalous couplings that can be
obtained at TESLA and LHC. To obtain the values for LHC we rescale the data
from the literature to a $b$ tagging efficiency of $50\%$ and keep the average
mistagging rate used of $1\%$ for other jets, which is somewhat optimistic. The
best LHC limits on $Vtc$ couplings come from top decays, whereas the best ones
on $Vtu$ couplings are from single top production.  The LHC limit on
$\mathrm{Br}(t \to Zc)$ with $\sigma_{\mu \nu}$ couplings is estimated to be
similar to the one with $\gamma_\mu$ couplings, since the same holds in our
analysis for TESLA. We assume one year of running time in all the cases, that
is, 100 fb$^{-1}$ for LHC, 300 fb$^{-1}$ for TESLA at 500 GeV and 500 fb$^{-1}$
for TESLA at 800 GeV. We use the statistical estimators explained before.

\begin{table}[htb]
\begin{center}
\begin{tabular}{lcccc}
& \multicolumn{2}{c}{LHC} & \multicolumn{2}{c}{TESLA} \\
& $95\%$ & $3\,\sigma$ & $95\%$ & $3\,\sigma$ \\
$\mathrm{Br}(t \to Zu)$ $(\gamma_\mu)$ &
  $6.2 \times 10^{-5}$ & $8.0 \times 10^{-5}$ &
  $1.9 \times 10^{-4}$ & $2.2 \times 10^{-4}$ \\
$\mathrm{Br}(t \to Zc)$ $(\gamma_\mu)$ &
  $7.1 \times 10^{-5}$ & $1.0 \times 10^{-4}$ &
  $1.9 \times 10^{-4}$ & $2.2 \times 10^{-4}$ \\
$\mathrm{Br}(t \to Zu)$ $(\sigma_{\mu \nu})$ &
  $1.8 \times 10^{-5}$ & $2.3 \times 10^{-5}$ &
  $6.2 \times 10^{-6}$ & $7.0 \times 10^{-6}$ \\
$\mathrm{Br}(t \to Zc)$ $(\sigma_{\mu \nu})$ &
  $7.1 \times 10^{-5}$ & $1.0 \times 10^{-4}$ &
  $6.2 \times 10^{-6}$ & $7.0 \times 10^{-6}$ \\
$\mathrm{Br}(t \to \gamma u)$ &
  $2.3 \times 10^{-6}$ & $3.0 \times 10^{-6}$ &
  $3.7 \times 10^{-6}$ & $3.6 \times 10^{-6}$ \\
$\mathrm{Br}(t \to \gamma c)$ &
  $7.7 \times 10^{-6}$ & $1.2 \times 10^{-5}$ &
  $3.7 \times 10^{-6}$ & $3.6 \times 10^{-6}$ \\
\end{tabular}
\end{center}
\caption{Best limits on top FCN couplings that can be obtained at LHC and TESLA
after one year of operation.
\label{tab:lim}}
\end{table}

We see that LHC and TESLA complement each other in the search for top FCN
vertices. The $\gamma_\mu$ couplings to the $Z$ boson can be best measured or
bound at LHC, whereas the sensitivity to the $\sigma_{\mu \nu}$ ones is better
at TESLA. For photon vertices, LHC is better for $\gamma tu$ and TESLA for
$\gamma tc$. The complementarity of LHC and TESLA also stems from the fact that
TESLA will not be able to distinguish $Ztq$ and $\gamma tq$ couplings in the
limit of its sensitivity, whereas LHC will because final states are different
and distinguish between them. On the other hand, the good charm tagging
efficiency expected at TESLA will allow to distinguish $Vtu$ and $Vtc$
couplings looking at the flavour of the final state jet, what is more difficult
to do at LHC.

\vspace{1cm}
\noindent
{\Large \bf Acknowledgements}

\vspace{0.4cm} \noindent
I am indebted to T. Riemann for many useful discussions. I also thank F. del
Aguila and A. Werthenbach for a critical reading of the manuscript and S.
Slabospitsky for
useful comments. I thank the members of the Theory group of DESY Zeuthen
for their warm hospitality during the realization of this
work. This work has been supported by a DAAD scholarship and
by the European Union under contract HTRN--CT--2000--00149

\end{document}